\documentclass[12pt]{article}
\newcommand{\be}{\begin{equation}}
\newcommand{\ee}{\end{equation}}
\newcommand{\ba}{\begin{eqnarray}}
\newcommand{\ea}{\end{eqnarray}}

\begin{document}
\title          {Domain Walls\\
            and Dimensional Reduction}
\author{C.D.\ Fosco\thanks{fosco@cab.cnea.gov.ar}\\
and\\
R.C. Trinchero\thanks{trinchero@cab.cnea.gov.ar}
\\
\\
{\normalsize\it
Centro At\'omico Bariloche,
8400 Bariloche, Argentina}}
\maketitle
\begin{abstract}
We study some properties of a dimensional reduction  mechanism for
fermions in an odd number $D+1$ of spacetime dimensions. A fermionic
field is equipped with a mass term with domain wall like defects along
one of the spacelike dimensions, which is moreover compactified. We
show that there is a regime such that the only relevant degrees of
freedom are massless fermionic fields in $D$ dimensions. For any
fixed gauge field configuration, the extra modes may be decoupled,
since they can be made arbitrarily heavy. This decoupling combines
the usual Kaluza-Klein one, due to the compactification, with a mass
enhancement for the non-zero modes provided by the domain wall
mechanism. We obtain quantitative results on the contribution of
the massive modes in the cases $D=2$ and $D=4$.
\end{abstract}

\bigskip

\section{Introduction.}
Domain wall fermions have been a very active subject of
theoretical research because of their many interesting properties and
applications.
They have recently attracted increased attention in their application
to Kaplan's~\cite{kap} domain wall approach to the problem of putting
chiral fermions on a lattice. This proposal has evolved into another,
more abstract construction, the so-called `overlap'
formalism~\cite{nar,overl,kap2}, that was designed in order to bypass
Nielsen-Ninomiya's no-go theorem~\cite{kars}.
It gives a procedure to
 define a chiral determinant in $D = 2 k$ dimensions in
terms of the `overlap' (that is, scalar product) between two different
Dirac vacua. These vacua correspond to two Dirac Hamiltonians in
$D= 2 k +1$ spacetime dimensions, in the presence of the same
{\em static\/} external gauge field, but with opposite signs for their
(Dirac) mass terms. This idea has been extensively applied to different
models, testing its predictions and consequences for both its continuum and
lattice versions. Particular interest has been directed to the issue of chiral
anomalies. The original idea has been also extended to the odd
dimensional case~\cite{odd}, and to the bosonic case~\cite{fosch}.
Recently, fermions of domain-wall like defects have also been studied
in the context of condensed matter systems~\cite{fl,ffl}.

Up to know, all the tests performed so far suggest that,
when properly regularized, domain wall fermions may indeed provide
a satisfactory definition of a chiral fermionic determinant.
This definition, either in its overlap or domain-wall version,
demands the introduction of a parameter $\Lambda$, with the dimensions
of a mass, which has to go to infinity at some point of the
construction. A finite-$\Lambda$ definition, on the other hand,
provides an effective definition of the chiral determinant, valid when
low-momentum external gauge-field configurations are considered.
Chiral $2 k$ dimensional modes are `confined' to a strip of width
$\sim \frac{1}{\Lambda}$ around a $2 k$-dimensional hypersurface defined
by the position of the domain wall, say, $x_{2k+1} = 0$.

In this paper we study domain wall fermions from the dimensional reduction
point of view. That is to say, we assume the extra dimension $x_{2k+1}$ to be
compactified, and we want to understand the combined effects of the domain
walls and the compact dimension on the effective dimensionally reduced theory.
The effective low-energy theory arising in this kind of system involves
fields of both chiralities, since the use of a compactified extra dimension
automatically produces a domain anti-domain wall pair and their companion
fermionic modes of opposite chiralities.
 We obtain the conditions that have to be satisfied by the parameters of the
theory, in order to  have a regime where the massive modes are decoupled.
In other words, we find the conditions for the theory to be effectively
dimensionally reduced to a model of massless Dirac fermions in $2k$ dimensions.

This paper is organized as follows: In section 2 we define and
discuss, in the functional integral approach, the dynamics corresponding
to the fermionic modes in an odd dimensional theory with a compactified
dimension, regarding the gauge field as external (non-dynamical).
We first deal with a restricted class of external gauge field configurations,
which are in fact similar to the static ones of the overlap formalism, and
afterwards extend the study to more general gauge field configurations.
We then show how to extract the effective dynamics of the reduced theory, and
to assure the decoupling of the higher modes. As explicit examples, the
decoupling for the Abelian case in $4+1$ and $2+1$ dimensions is discussed
in the framework of perturbation theory.

\section{Dimensional reduction}
The Euclidean functional integral corresponding to a fermionic
field in the presence of an external gauge field $A$, in $D+1$
dimensions, and with a mass term depending on the extra coordinate
$x_{D+1}\equiv s$, is defined by
\be
{\cal Z} (A) \;=\; \int \, [d{\bar \Psi}] [d\Psi] \,
\exp \left(- \int d^{D+1}x \, {\bar \Psi} {\cal D} \Psi \right)
\label{defza}
\ee
where the Euclidean Dirac operator ${\cal D}$ appearing in the
fermionic action is
\be
{\cal D} \;=\; \gamma_s D_s + m(s) + \not \!\! D (A) \,.
\label{defcd}
\ee
We shall adopt the conventions that Greek indices run from $1$ to $D$;
$x=(x_\mu), \mu = 1 \cdots D$ will denote the $D$ uncompactified
`physical' dimensions, while $s$ stands for the compactified
one ($x_{D+1}$). Also, Dirac's $\gamma$ matrices are Hermitian,
obeying the relations
\be
\{ \gamma_\mu , \gamma_\nu \} \;=\; 2 \delta_{\mu\nu} \;\;,\;\;
\{ \gamma_\mu , \gamma_s \} \;=\; 0 \;\;, \gamma_s^2 = 1 \;.
\ee

\subsection{Restricted gauge field configurations}\label{rest}
We want to express the Euclidean functional integral (\ref{defza}) for
${\cal Z}(A)$ in a form that renders the existence and properties of the
domain wall fermions more transparent. To this effect we shall
introduce an expansion of the fermionic fields in terms of a suitable
Hermitian operator.
As ${\cal D}$ is not Hermitian, we introduce the explicitly Hermitian,
positive definite operator ${\cal H}$, defined in terms of ${\cal D}$ as
follows:
\be
{\cal H} \;=\; {\cal D}^\dagger {\cal D} \;.
\label{defh}
\ee
We first assume that $A_{D+1} = 0$, and $A_\mu=A_\mu(x)$. This
restriction (to be relaxed in subsection \ref{relax}) is imposed
in order to have a clean factorization into the dynamics along the
$x$-coordinates and along $s$, as then  ${\cal H}$ is the sum of two 
commuting pieces
\be
{\cal H} \;=\; h \,- \, {\not \!\!D}^2
\ee
where $h$ acts on the extra coordinate $s$ only
\be
h \;=\; - \partial_s^2  + m^2 (s) - \gamma_s m'(s) \;,
\ee
with $m'(s) = \frac{d}{ds}m(s)$.
We also define projectors corresponding to the two eigenvalues
$\pm 1$ of $\gamma_s$, named ${\cal P}_{L,R}$, respectively, and
two operators $a, a^\dagger$, acting on the Hilbert space of functions
depending on the coordinate $s$
\be
a\;=\; \partial_s + m(s) \;\;,\;\; a^\dagger \;=\; - \partial_s + m(s)
\;.
\ee
The boundary conditions for the functions upon which these operators
act are always going to be such that they become relatively adjoint.
That will hold true for both the compact and non-compact cases.

After these definitions, ${\cal H}$ may be written as
\be
{\cal H} \;=\; a^\dagger a {\cal P}_L + a a^\dagger {\cal P}_R
- {\not \!\! D}^2
\label{haa}
\ee
or
\be
{\cal H} \;=\; (a^\dagger a - {\not \!\! D}^2){\cal P}_L +
               (a a^\dagger - {\not \!\! D}^2){\cal P}_R \;.
\ee
We define normalized eigenfunctions corresponding to the Hermitian
and positive operators $a^\dagger a$ and $a a^\dagger$,
\ba
a^\dagger a f^{(+)}_n (s) &=& {\cal E}^2_n f^{(+)}_n (s) \nonumber\\
a a^\dagger f^{(-)}_n (s) &=& {\cal E}^2_n f^{(-)}_n (s) \;,
\ea
and note that, as the extra dimension is compactified, both
operators shall have zero modes, namely, eigenfunctions
$f^{(+)}_0 (s)$ and $f^{(-)}_0 (s)$
with  ${\cal E}_0 = 0$. They shall correspond to the equations
\be
a f^{(+)}_0 (s) = 0 \;\;,\;\;a^\dagger f^{(-)}_0 (s) = 0 \;,
\ee
respectively. Had the extra dimension been of infinite extension,
one of the zero modes $f^{(+)}$, $f^{(-)}$ would have disappeared,
because it must have had an infinite norm. We shall here assume
that the extra dimension is finite: $s \in [-L,L]$, and that $m(s)$
is periodic, so that we always have both zero modes. Moreover, they can
be explicitly written as
\be
f^{(\pm)}_0 (s) \;=\; N^{(\pm)}_0 \exp[\mp \int_0^s dt \, m(t)]
\ee
where
\be
N^{(\pm)} \;=\; \left\{
\int_{-L}^{+L} ds \exp[ \mp \,2 \int_0^s dt m(t)]
\right\}^{-\frac{1}{2}} \;.
\ee
It is convenient at this point to be more specific about the
functional form of $m(s)$. The simplest possibility within
the compactified case is to give $m(s)$ a positive step
at $s=0$
\be
m(s) \;=\; \Lambda \; {\rm sign} (s) \;,
\ee
which, because of periodicity, does also have a negative jump
at $s = +L \equiv -L$. This yields for the zero modes the
explicit expression
\ba
f_0^{(+)}(s) &=& \sqrt{\frac{\Lambda}{1-e^{-2 \Lambda L}}} \;
e^{- \Lambda |s|} \nonumber\\
f_0^{(-)}(s)&=& \sqrt{\frac{\Lambda}{e^{2 \Lambda L}-1}} \;
e^{\Lambda |s|} \;,
\ea
which shall correspond to modes localized around $s=0$ and $s=L$,
respectively.
In the uncompactified case, on the other hand,  as $L \to \infty$,
only the mode $f_0^{(+)}$ survives. It is evident how to extend
the previous zero mode solutions to the general case of a
stepwise mass profile $m(s)$ such that $m^2(s)=\Lambda^2$.

Keeping (\ref{haa}) in mind, we see that the eigenfunctions of
${\cal H}$ shall be products of eigenfunctions of
$a^\dagger a {\cal P}_L$ or $a a^\dagger {\cal P}_R$, times
eigenfunctions of $-{\not\!\!D}^2$.
We then expand the fermionic fields $\Psi (x,s)$ and ${\bar \Psi}(x,s)$
in terms of these eigenfunctions. Of course, each term
will also depend on an arbitrary $D$-dimensional spinor $\psi^{(n)}(x)$:
\ba
\Psi (x,s) &=& \sum_n \left[ f^{(+)}_n (s) {\cal P}_L \psi^{(n)}(x)
+ f^{(-)}_n (s) {\cal P}_R \psi^{(n)}(x)
\right] \nonumber\\
{\bar \Psi} (x,s) &=& \sum_n \left[ f^{(+)\dagger}_n (s)
{\bar \psi}^{(n)}(x){\cal P}_R + f^{(-)\dagger}_n (s)
{\bar \psi}^{(n)}(x){\cal P}_L
\right] \;,
\label{delpr}
\ea
or
\ba
\Psi (x,s) &=& \sum_n \left[ f^{(+)}_n (s) \psi_L^{(n)}(x)
+ f^{(-)}_n (s)  \psi_R^{(n)}(x)
\right] \nonumber\\
{\bar \Psi} (x,s) &=& \sum_n \left[
{\bar \psi}_L^{(n)}(x)f^{(+)\dagger}_n (s)  +
{\bar \psi}_R^{(n)}(x) f^{(-)\dagger}_n (s)
\right] \;,
\label{delr}
\ea
with $\psi_{L,R}^{(n)}(x) = {\cal P}_{L,R} \psi^{(n)} (x)$, and
${\bar \psi}_{L,R}^{(n)}(x) = {\bar \psi}^{(n)}(x){\cal P}_{R,L}$.

We could, of course, expand each $\psi^{(n)}$ in terms of the
eigenfunctions of $-\not \!\!D^2$, but that shall not be necessary.
Note that no approximation has been invoked in order to obtain
the expansions (\ref{delpr}) and (\ref{delr}), since we are only
applying the property that ${\cal H}$ can be written as the sum
of two commuting pieces, and we can always expand any state in
such a way as to make one of the pieces diagonal.

Using expansion (\ref{delr}), we have a decomposition of the fermionic
integration measure,
\be
[d  \Psi] \; [d {\bar \Psi}] \;=\; \prod_n \left(
[d \psi^{(n)}_L][d {\bar \psi}^{(n)}_L]
[d \psi^{(n)}_R][d {\bar \psi}^{(n)}_R]
\right) \;,
\label{deme}
\ee
and a series for the fermionic action. It is convenient, in this
series, to separate what corresponds to the zero modes, from
the contributions due to the higher $n$'s, since they shall have
quite different properties. Explicitly, it reads
$$
S \;=\; \int d^D x [ {\bar \psi}^{(0)}_L (x) \not \!\!
D \psi^{(0)}_L (x) \;+\;{\bar \psi}^{(0)}_R (x)
\not \!\! D \psi^{(0)}_R (x)  ]
$$
$$
\sum_{n\neq 0} \int d^D x \left\{
{\bar \psi}^{(n)}_L (x) \not \!\!
D \psi^{(n)}_L (x) \;+\;{\bar \psi}^{(n)}_R (x)
\not \!\! D \psi^{(n)}_R (x) \;+\;
\right.
$$
\be
\left.
{\cal E}_n [
{\bar \psi}^{(n)}_R (x) \psi^{(n)}_L (x) +
{\bar \psi}^{(n)}_L (x) \psi^{(n)}_R (x) ]
\right\} \;.
\label{deac}
\ee

This shows  that there appear massless $D$-dimensional fermionic
fields localized around the domain wall defects, and that there
is an infinite tower of massive states. These states can be decoupled
of the massless ones, since their masses ${\cal E}_n$ will satisfy,
as shown below, the inequality
\be
{\cal E }_n \;\geq\; \sqrt{|\Lambda|^2 + (\frac{\pi}{L})^2} \;.
\label{ineq}
\ee
Thus the masses of these higher modes can be made arbitrarily
large by proper choices of the parameters $\Lambda$, $L$.

Let us proof inequality (\ref{ineq}) for the eigenvalues of $h$
corresponding to the modes with $n \neq 0$. We first use the
property that, for stepwise profiles $m(s)$, such that
$m^2 (s) = \Lambda^2$, the only localized
states are the zero modes. This can be proved just by taking into
account that localized states are combinations of
real exponentials, and those combinations are completely determined once
the matching conditions due to the $\delta$ functions are
imposed. For the non-zero modes, the states are combinations of
exponentials of $\pm i k \cdot x$. The periodicity of the extra coordinate
fixes the minimum $k$ to be $\frac{\pi}{L}$, and the $\delta$-function
term only fixes boundary conditions, thus for any
eigenstate $f^{(\pm)}_n$,
\be
\langle f^{(\pm)}_n|h|f^{(\pm)}_n\rangle \;=\; {\cal E}_n^2 \;
\geq \; (\frac{\pi}{L})^2 + \Lambda^2 \;.
\ee

Let us consider the issue of decoupling. Equations (\ref{deme}) and
(\ref{deac}) already indicate that
\be
{\cal Z} \;=\; {\cal Z}^{(0)} \,\times \, \prod_{n \neq 0}
{\cal Z}^{(n)}
\ee
where ${\cal Z}^{(0)}$ contains the massless modes of the dimensionally
reduced theory
$$
{\cal Z}^{(0)} \,=\,\int [d {\bar \psi}^{(0)}_L] [d \psi^{(0)}_L]
[d {\bar \psi}^{(0)}_R] [d \psi^{(0)}_R]
$$
\be
\times \,\exp \left[ - \int d^Dx ( {\bar \psi}^{(0)}_L \not \!\! D
\psi^{(0)}_L \,+\, {\bar \psi}^{(0)}_R \not \!\! D \psi^{(0)}_R )
\right]\,, 
\label{dom} 
\ee 
while 
\ba 
{\cal Z}^{(n)} &=& \int [d
{\bar \psi}^{(n)}] [d \psi^{(n)}] \exp \left[ - \int d^Dx 
( {\bar\psi}^{(n)}\not \!\! D \psi^{(n)} \,+\, {\cal E}_n {\bar
\psi}^{(n)}\psi^{(n)} )\right]\nonumber\\ &=& \det ( \not \!\! D +
{\cal E}_n )
\label{dets}
\ea
are the massive, decoupling modes.
Of course they will only decouple if some conditions are imposed
on the external gauge fields, namely, their momentum dependence
cannot be arbitrary. In order to achieve decoupling, we have to
assume that the external momenta $p_\mu$, corresponding to the
Fourier components of the gauge fields, are small in comparison to
${\cal E}_n$. This guarantees the existence of a momentum
expansion for the massive determinants of (\ref{dets}). The terms
of such an expansion are suppressed by powers of $\frac{p^2}{{\cal
E}_n^2}$, since the Feynman diagrams are analytic below the
threshold.
\subsection{General gauge field configurations}\label{relax}
In the preceding subsections we have addressed the issue of 
decoupling in the presence of external gauge fields satisfying the 
conditions, 
\begin{equation}\label{2.3.1}
\partial_s A_{\mu}=0 \;\;, \qquad A_s=0 \;.
\end{equation}

As mentioned in subsection \ref{rest}, under conditions (\ref{2.3.1}),
the eigenfunctions of ${\cal D}^\dagger {\cal D}$  can be factored into 
eigenfunctions of $h$ times eigenfunctions of $(\not\!\!D)^2$. In the present subsection 
we present a perturbative proof of the fact that, even for the general 
case, namely, relaxing (\ref{2.3.1}), we still have a non-vanishing gap 
of ${\cal{O}}( \Lambda )$ in the spectrum of ${\cal D}^\dagger {\cal D}$. 
The spirit of the proof is to use the fact that, by performing a
perturbative expansion around the restricted case (\ref{2.3.1}),
the correction to the gap is negligible when the gauge field is smooth. 

We split  ${\cal D}^\dagger {\cal D}$ into free (${\cal{H}}$) and 
perturbation ($V$) terms                                              
\begin{equation}\label{2.3.2}
{\cal{D}}^{\dagger}{\cal{D}}= {\cal{H}} + V
\end{equation}
where ${\cal{H}}$ is the one discussed in \ref{rest}, i.e.,
\begin{equation}\label{2.3.3}
{\cal{H}} = m(s)^2 - \gamma_s m^{'}(s) - \partial_s^2 -
(\not\!\!D)^2 (A_{\mu}(x, s=0))
\end{equation}
and $V$ is defined by
\begin{equation}\label{2.3.4}
V\;=\; - \partial_s A_s - A_s^2 - \gamma_{\mu} \gamma_s  F_{\mu s} +
\not\!\!{D}^2 (A_{\mu}(x, s=0)) - \not\!\!{D}^2 (A_{\mu}) \;.
\end{equation}

The free term is 
${\cal{O}}( \Lambda^2 )$, while for the perturbation we shall assume
that the gauge field is smooth, in the same sense as in the restricted 
case, namely, their derivatives are small when compared with $\Lambda$.
Thus the perturbation is ${\cal{O}}( \Lambda^0 )$. 
It is also useful to note that the perturbation is a periodic function 
of $s$. 

In order to compute the corrections to the spectrum of (\ref{2.3.3})
we should consider matrix elements of the following kind:
\begin{equation} \label{2.3.5}
<n|V|m> = \int_{-L}^{L} ds \;\; f_n^{\pm\dagger}(s) \left\{ \int
d^d x
 {\bar{\psi}}_{ L R  }^n (x)
V(x,s) \psi_{ LR }^n (x)  \right\} f_n^{\pm}(s)
\end{equation}
where $V(x,s)$ may be written as a Fourier series,
\begin{equation} \label{2.3.6}
V(x,s) = \sum_n \left[ V_n^e (x) cos(\frac{n \pi s}{L})  + V_n^o (x)
sin(\frac{n \pi s}{L})\right]
\end{equation}
and the functions $f_n^{\pm}(s)$ appearing in (\ref{2.3.5}) are the
excited states eigenfunctions of $h$. They are odd or even  ($h$
is invariant under $s \leftrightarrow -s$) and are explicitly 
given by,
\begin{eqnarray}
f_n^{+,o}(s)   & = & \frac{1}{\sqrt{L}} sin(\frac{n \pi s}{L}) =
f_n^{-}(s) \;\;, \qquad 
\nonumber\\ f_n^{+,e}(s)   & = & \frac{1}{\sqrt{ L ( 1 + (
\frac{\pi L \Lambda}{n} )^2)}} \left( \cos( \frac{n \pi s}{L}) \mp
\frac{a \Lambda}{n \pi} \sin(\frac{n \pi s}{L}) \right)
\label{2.3.7}
\end{eqnarray}
with eigenvalues,
\begin{equation} \label{2.3.8}
\epsilon_n ^2= (\frac{n \pi }{L})^2 + \Lambda^2
\end{equation}
in both cases. In order to estimate the perturbative correction,
we see that the $\Lambda$ dependence of the matrix elements 
(\ref{2.3.5}) is determined by the $s$-integration, since the parameter
$\Lambda$ only appears in the $f_n^{\pm}$ functions. 
From formulae (\ref{2.3.6}) and (\ref{2.3.7}) it is evident that this
matrix elements are at most of ${\cal O}( \Lambda^0 )$. Therefore,
when the Fourier components of the gauge fields are
small compared with $\Lambda$, the corrections to the unperturbed
energy levels (\ref{2.3.8}) do not eliminate the gap of 
${\cal O}( \Lambda )$  of the unperturbed theory.  Indeed, 
the corrections to the eigenvalues are not capable of modifying a
gap of order $\Lambda$, since they always involve  
${\cal O}( \Lambda^0 )$ terms.
                                                   
\subsection{The fermionic propagator}
The results of 2.1 were important in order to understand the mechanism
on decoupling, from the point of view of the gauge field action.
Namely, the contribution to the effective gauge field action due
to the fermion loops is shown to be given mainly from the contribution
of the massless, domain wall modes. However, for physical processes
involving external fermions, we need to consider a different case,
since we have to include fermionic sources ${\bar \eta} (x,s)$
$\eta (x,s)$ into the functional integral (\ref{defza}):
$$
{\cal Z}({\bar \eta},\eta;A)\;=\; [{\cal Z}(A)]^{-1} \int
[d {\bar \Psi}] [d {\Psi}] \; \exp \left\{ - S \right.
$$
\be
\left. +\; \int ds d^Dx [ {\bar \eta}(x,s) \Psi (x,s) + {\bar \Psi} (x,s)
\eta (x,s) ] \right\} \;.
\ee
Expanding the measure and the action, according to (\ref{deme})
and (\ref{deac}), respectively, we have, after integrating out the
fermions
\be
{\cal Z}({\bar \eta},\eta;A)\;=\;
{\cal Z}^{(0)}({\bar \eta},\eta;A) \times  \prod_{n \neq 0}
{\cal Z}^{(n)}({\bar \eta},\eta;A) \;.
\ee
${\cal Z}^{(0)}({\bar \eta},\eta;A)$ is the generating
functional corresponding to the domain wall fermions:
$$
{\cal Z}^{(0)}({\bar \eta},\eta;A) \;=\;
$$
$$
\exp \left\{- \int ds d^Dx \int ds'd^D x' \,
{\bar \eta} (x,s) [ f_0^{(+)\dagger}(s) {\cal P}_R +f_0^{(-)\dagger}(s)
{\cal P}_L]
\right.
$$
\be
\left. \not \!\! D^{-1} (x,x') \,
[f_0^{(+)}(s') {\cal P}_L + f_0^{(-)}(s'){\cal P}_R]
\eta (x',s') \right\} \;,
\label{zfe0}
\ee
while ${\cal Z}^{(n)}({\bar \eta},\eta;A)$ contains the infinite
tower of massive fermionic fields
$$
{\cal Z}^{(n)}({\bar \eta},\eta;A)\;=\;
$$
$$
\exp \left\{- \int ds d^Dx \int ds'd^D x' \,
{\bar \eta} (x,s) [ f_0^{(+)}(s) {\cal P}_R +f_0^{(-)}(s){\cal P}_L]
\right.
$$
\be
\left. (\not \!\! D + {\cal E}_n)^{-1} (x,x') \,
[f_0^{(+)}(s') {\cal P}_L + f_0^{(-)}(s'){\cal P}_R]
\eta (x',s')\right. \;.
\label{zfe1}
\ee
Again, decoupling is achieved when the masses (${\cal E}_n$) in
the propagators derived from (\ref{zfe1}) are large compared with the
external momenta (now the momenta of the fermions are also relevant).
The domain wall piece (\ref{zfe0}) allows us to write
down their contribution to the fermionic propagator
$$
\langle \Psi (x,s) {\bar \Psi}(x',s') \rangle \;=\;
$$
\be
\frac{\Lambda}{1-e^{-2 \Lambda L}} e^{- \Lambda (|s|+|s'|)}
\not \!\! D^{-1}(x,x') {\cal P}_L \;+\;
\frac{\Lambda}{e^{2 \Lambda L}-1} e^{\Lambda (|s|+|s'|)}
\not \!\! D^{-1}(x,x') {\cal P}_R \;.
\ee
It is straightforward to obtain the expression for $j_\mu (x,s)$,  the
vacuum current contribution due to this part of the fermion propagator.
With the usual definitions, we see that
$$
j_\mu (x,s)\;=\; \langle {\bar\Psi} (x,s) \gamma_\mu \Psi (x,s) \rangle
$$
$$
- \, {\rm tr}
\left[\gamma_\mu \langle\Psi (x,s){\bar \Psi}(x,s)\rangle\right]
$$
\be
\frac{\Lambda}{1-e^{-2 \Lambda L}} e^{-2 \Lambda |s|}
j^L_\mu (x)
\,+\,\frac{\Lambda}{e^{2 \Lambda L}-1}  e^{2 \Lambda |s|}
j^R_\mu (x)
\ee
where $j^L_\mu (x)$ and $j^R_\mu (x)$ denote the corresponding vacuum
currents in $D$ dimensions:
\ba
j^L_\mu (x)&=&\langle {\bar\psi}_L (x)\gamma_\mu\psi_L (x)\rangle\nonumber\\
j^R_\mu (x)&=&\langle {\bar\psi}_R (x)\gamma_\mu\psi_R (x)\rangle \;.
\ea
It becomes clear from the above that the contributions of the chiral zero
modes to the current are localized around each domain wall, with a
localization length $\sim \frac{1}{\Lambda}$. In order to be able to
resolve the two currents, the localization length should be smaller than
the compactification length, what amounts to the inequality
\be
\Lambda \times L \; > \; 1 \;.
\ee
Regardless of whether the two chiral currents have a large overlap or not,
the integral of the $D+1$ dimensional current $j_\mu(x,s)$ along the $s$
coordinate always produces the result corresponding to the current of a
Dirac fermions in $D$ dimensions
\be
\int_{-L}^L \,ds \,j_\mu (x,s) \;=\; j_\mu (x) \;=\;
\langle{\bar\psi}(x)\gamma_\mu \psi(x)\rangle \;,
\ee
where $\psi \,=\, \psi_L + \psi_R$.

We shall also present, for the sake of completeness, a derivation of
the fermionic propagator obtained by directly inverting the operator
${\cal D}$ of (\ref{defza}), in $4+1$ dimensions. We consider the free
($A=0$) case for the sake of clarity, since, for this calculation, the
necessary changes for the non-free case are easy to introduce. Moreover,
this free propagator in $4+1$ is used in the next subsection, in the
perturbation theory example.
This system is described by the free Euclidean action:
\be
S \;=\; \int ds d^D x \, {\cal L}_F \;\;,\;\;
{\cal L}_F \;=\; {\bar \Psi} ( \not \! \partial + m(s) )
\Psi
\label{fract}
\ee
The free Dirac operator $\not \! \partial$ acts on the five co-ordinates:
\be
\not\! \partial \;\equiv\; \gamma_I \partial_I \;=\;
\gamma_s \partial_s\,+\,\gamma_\mu \partial_\mu \;.
\ee

Dirac's matrices are chosen to be in the representation:
\be
\gamma_\mu \;=\; \left(\begin{array}{cc}
0 & \sigma^\dagger_\mu \\
\sigma_\mu & 0
\end{array}
\right) \;\; , \;\;
\gamma_s \;=\;\left(\begin{array}{cc}
1 & 0 \\
0 & -1
\end{array}
\right)
\;\;,\;\; \sigma_\mu \,=\, ({\vec \sigma}, i 1)
\ee
where ${\vec \sigma}$ denotes the three familiar Pauli's matrices.
Finally, the domain wall mass term $m(s)$ is assumed to
be of the form:
\be
m(s) \;=\; \Lambda \, {\rm sign} (s)
\label{wall}
\ee
namely, it contains a domain wall like effect localized at $s=0$,
and an anti domain wall companion at $s=L$.
>From its definition, it follows that the fermionic propagator
$S$ satisfies
\be
[ \not \! \partial_{x,s} + m(s) ] S(x,s;x',s') \;=\;
\delta^{(4)} (x-x') \delta (s-s')
\label{eqpro}
\ee
where the subscripts $x,s$ mean that derivatives act with respect
to the $x,s$-co-ordinates. Because of translation invariance in
$x_\mu$, a Fourier transformation in these co-ordinates suggests
itself:
\be
S(x,s;x',s') \;=\; \int \frac{d^4 k}{(2 \pi)^4} \,
e^{i k_\mu (x-x')_\mu}\, {\tilde S}_k (s,s')\;.
\ee
Hence equation (\ref{eqpro}) implies for ${\tilde S}_k$
\be
[\gamma_s \partial_s + i \gamma_\mu k_\mu] {\tilde S}_k
(s,s') \;=\;\delta (s - s') \;.
\ee
To solve this equation, it is convenient to define an
auxiliary function $G_k$, determined from $S_k$ by
the relation
\be
{\tilde S}_k (s,s') \;=\; [-\gamma_s \partial_s - i
\gamma_\mu k_\mu + m(s) ] G_k (s,s')
\label{act}
\ee
which substituted into (\ref{eqpro}) yields for $G_k$ the
equation
\be
[\partial_s^2 + 2 \Lambda \gamma_s (\delta (s) - \delta (s-L)) +
k^2 + \Lambda^2] G_k (s,s') \;=\; \delta (s,s')
\ee
where $k^2 = k_\mu k_\mu$.
Then, $G_k$ is the inverse of a Hamiltonian operator
${\cal H}_k$ which contains a delta-like potential in a one-dimensional
quantum mechanical system
\be
{\cal H}_k \;=\; - \frac{d^2}{d s^2} +
2 \Lambda \gamma_s (\delta (s) - \delta (s-L)) + \omega_k^2
\ee
where $\omega_k^2 = k^2 + \Lambda^2$.
In order to invert this operator, it is convenient to diagonalize it,
and then we are back into the issue of knowing the spectrum of $h$,
the operator considered in the general derivation of the previous
subsection. Again, the domain wall modes correspond to the
bound states of the Hamiltonian. Keeping only the contribution
coming from these states, we get for $G_k$
$$
G_k(s,s')\;=\; \frac{\Lambda}{k^2} \left[ (e^{2\Lambda L}-1)^{-1}
e^{\Lambda(|s|+|s'|)} {\cal P}_L \right.
$$
\be
\left. + (1- e^{-2\Lambda L})^{-1}
e^{-\Lambda(|s|+|s'|)} {\cal P}_R \right]
\ee
which, when introduced in (\ref{act}) yields,
$$
{\tilde S}_k (s,s') \;=\;
\frac{\Lambda}{1-e^{-2 \Lambda L}} e^{- \Lambda (|s|+|s'|)}
(i \not \! k )^{-1} {\cal P}_L
$$
\be
\frac{\Lambda}{e^{2 \Lambda L}-1} e^{\Lambda (|s|+|s'|)}
(i \not \! k )^{-1} {\cal P}_R \;.
\ee

\subsection{Decoupling of the massive modes}
To render the general derivations of the previous paragraphs more
concrete, we present here their realization for two particular examples.
We consider the cases $D=4$ and $D=2$, with Abelian external gauge fields.
Making the redefinitions $A \to i e A$, and
recalling equations (\ref{dom}) and (\ref{dets}), we may write
for the full functional ${\cal Z}(A)$,
\be
{\cal Z} (A) \;=\; \exp \left[ - \Gamma (A) \right]
\ee
where the effective action $\Gamma (A)$ is a sum
\be
\Gamma (A) \;=\; \Gamma^{(0)}(A) \;+\; \sum_{n\neq 0}
\Gamma^{(n)}(A)
\ee
with the domain wall effective action
\be
\Gamma^{(0)} \;=\; - {\rm Tr} \ln [\not\!\partial + i \not\!\!A]
\ee
and the contribution of the massive modes
\be
\Gamma^{(n)} \;=\; - {\rm Tr} \ln [\not\!\partial
+ i \not\!\!A + {\cal E}_n] \;.
\ee
Note that the issue of the convergence of the series over $n$
is not clear at all, unless one obtain a more explicit expression,
in terms of $A$, for the contribution
of the massive modes.
To check this in the $D=4$ case, we may use the known results~\cite{eulh}
for the
 derivative expansions of the Abelian fermionic determinants in
$4$ dimensions, obtaining for the leading terms
\be
\Gamma^{(n)}_{D=4} \,=\, \int d^4 x \left\{ -\frac{e^2}{120 \pi^2 {\cal
E}^2_n} \,F_{\mu\nu} \partial^2 F_{\mu\nu} + \frac{e^4}{1440 \pi^2 {\cal
E}^4_n} [(F^2)^2 + \frac{7}{16} (F {\tilde F})^2] \right\}\;,
\ee
while in $D=2$ a similar calculation yields
\be
\Gamma^{(n)}_{D=2} \,=\, \int d^2 x \left\{
\frac{e^2}{24 \pi {\cal E}^2_n} \,F_{\mu\nu}F_{\mu\nu} +
\frac{e^2}{360\pi {\cal E}^4_n}  F_{\mu\nu} \partial^2 F_{\mu\nu}\right\}\;,
\ee where we are of course neglecting higher dimensional terms.
Thus, in order to assure the decoupling of these unwanted
contributions to $\Gamma$, either in $D=4$ or $D=2$, we must be able to
render
\be
\sum_{n\neq 0} \frac{1}{{\cal E}^2_n} \;\;,\;\;
\sum_{n\neq 0} \frac{1}{{\cal E}^4_n} \;,
\ee
(and all the series with higher even powers of ${\cal E}_n$ in the
denominator) arbitrarily small by tuning $\Lambda$ and $L$.
But that can always be done, because for large $n$, the eigenvalues of
${\cal E}_n$ will obviously grow at least like $n$, what makes the series
above absolutely convergent. Then we adjust $\Lambda$ and $L$ to make
those sums arbitrarily small (note that we could interchange the order
of the limit $\Lambda \to \infty$ or $L \to 0$ and the summation of the
series). It goes without saying that higher dimensional terms carry higher
powers of $\frac{1}{{\cal E}_n}$, what makes them more convergent.

This concludes our particular examples of decoupling. It is easy to verify
that the convergence arguments remain the same for the case of external
non-Abelian fields, since the power of ${\cal E}_n$ that appears is 1
fixed by dimensional arguments, identical for the Abelian and non-Abelian
cases.

\subsection*{Acknowledgements} C.~D.~F. and R.~C.~T. are supported by 
Instituto Balseiro and CONICET, Argentina.
This work is  supported in part by grants from 
ANPCYT (PICT 97/0053, PICT 97/0358) and Fundaci\'on Antorchas,
Argentina. R.~C.~T. would like to thank the Abdus Salam ICTP for
financial support.


\begin{thebibliography}{99}
\bibitem{kap}D. B. Kaplan, Phys. Lett. {\bf B288} 342 (1992).\\
For a review of Kaplan's formulation, see: K. Jansen,\\
Phys. Rept. {\bf 273}: 1-54 (1996).
\bibitem{nar}R. Narayanan and H. Neuberger, Nucl.
Phys.{\bf B443}, 305 (1995). See also:
R. Narayanan, H. Neuberger and P. Vranas,
Nucl. Phys. Proc. Suppl.{\bf 47}: 596-598 (1996);
R. Narayanan and H. Neuberger, Nucl. Phys. Proc. Suppl.
{\bf 47}: 591-595 (1996);
H. Neuberger, hep-lat/9511001.
\bibitem{overl}S. Randjbar-Daemi and J. Strathdee, Nucl. Phys.
{\bf B443}:386-416 (1995);\\
S. Randjbar-Daemi and J. Strathdee, Nucl. Phys. {\bf B466}:
335-360 (1996).
\bibitem{kap2}D.B. Kaplan and M. Schmaltz, Phys. Lett. {\bf B368}, 
44 (1996).
\bibitem{kars} L. H. Karsten, Phys. Lett. {\bf 104B}, 315 (1981);\\
L. H. Karsten and J. Smit, Nucl. Phys {\bf B183}, 103 (1981);\\
H. B. Nielsen and M. Ninomiya,  Nucl. Phys. {\bf B185}, 20 (1981);
Nucl. Phys {\bf B193}, 173 (1981); Phys. Lett. {\bf 105B}, 219 (1981).
For a review, see: Y. Shamir, Nucl. Phys. Proc. Suppl.
{\bf 47}: 212-227 (1996).
\bibitem{odd}Y. Kikukawa and H. Neuberger,
e-Print Archive: Nucl. Phys. {\bf B513}, 735 (1998).
\bibitem{fosch}C. D. Fosco and F. A. Schaposnik;
Phys. Lett. {\bf B432}191 (1998). 
\bibitem{fl}C.~D.~Fosco and A.~L\'opez; Nucl. Phys. {\bf B538},
685 (1999). 
\bibitem{ffl}C.~D.~Fosco, E.~Fradkin and A.~L\'opez; 
Phys. Lett. {\bf B451}, 31 (1999). 
\bibitem{eulh}W. Heisenberg and H. Euler; Z. Phys. 98, 714 (1936).
\end{thebibliography}
\end{document}